\documentclass{ws-procs9x6}

\begin{document}

\title{STAR POLYMERS IN CORRELATED DISORDER}

\author{V. BLAVATS'KA}

\address{Institute for Condensed Matter Physics,
National Academy of Sciences of Ukraine\\
Lviv 79011, Ukraine \& \\
Institut f\"ur Theoretische Physik,
Universit\"at Leipzig\\
Leipzig 04109, Germany\\
E-mail: viktoria@icmp.lviv.ua}

\author{C. VON FERBER}

\address{Applied Mathematics Research Centre, Coventry University,\\
Coventry CV1 5FB, UK \& \\
Physikalisches Institut,
Universit\"at Freiburg\\
Freiburg 79104, Germany\\
E-mail: C.vonFerber@coventry.ac.uk}

\author{YU. HOLOVATCH}

\address{Institute for Condensed Matter Physics,
National Academy of Sciences of Ukraine\\
Lviv 79011, Ukraine \& \\
Institut f\"ur Theoretische Physik,
Johannes Kepler Universit\"at Linz\\
Linz 4040, Austria\\
E-mail: hol@icmp.lviv.ua}

\begin{abstract}
We analyze the impact of a porous medium (structural disorder) on
the scaling of the partition function of a star polymer immersed in a
good solvent. We show that corresponding scaling exponents change if
the disorder is long-range-correlated and calculate the exponents in
the new universality class. A notable finding is that star and
chain polymers react in  {\em
qualitatively} different manner on the presence of disorder: the corresponding scaling exponents
{\em increase} for chains and {\em decrease} for stars. We
discuss the physical consequences of this difference.
\end{abstract}


\bodymatter

\section{Introduction}\label{hol:sec1}

Polymer theory may serve as an archetype of an approach where the
application of the path integral formalism leads both to a
quantitative understanding of a whole range of physical, chemical,
and biological phenomena as well as to their accurate quantitative
description.\cite{desCloizeaux90,Kleinert95,Schaefer99} Most
directly this is shown by the Edwards model that describes a polymer
chain in terms of a path integral and takes into account chain
connectivity and self-avoiding interaction.\cite{Schaefer99} A
textbook derivation maps this simple two-parameter model to the
$m=0$ de~Gennes limit\cite{deGennes79} of the $O(m)$ symmetric field
theory. Standard field theoretical tools explain the origin of the
scaling laws that govern polymer structural behaviour and allow to
calculate the exponents that govern these scaling laws with high
accuracy.

\begin{figure}
\centerline{ \psfig{file=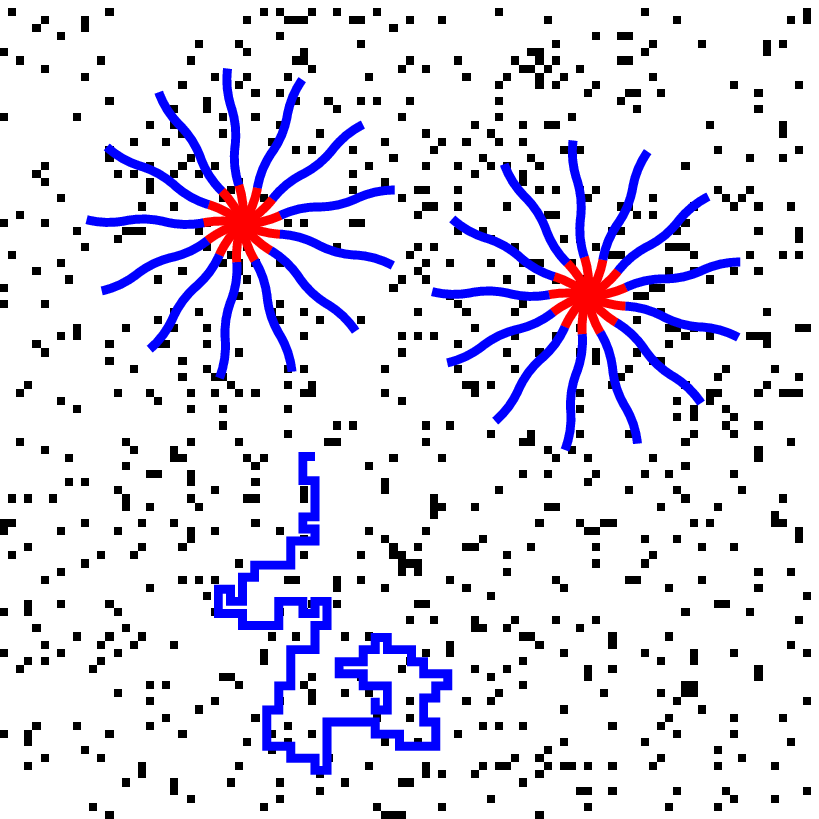,width=1.8in}
\hspace{0.5in} \psfig{file=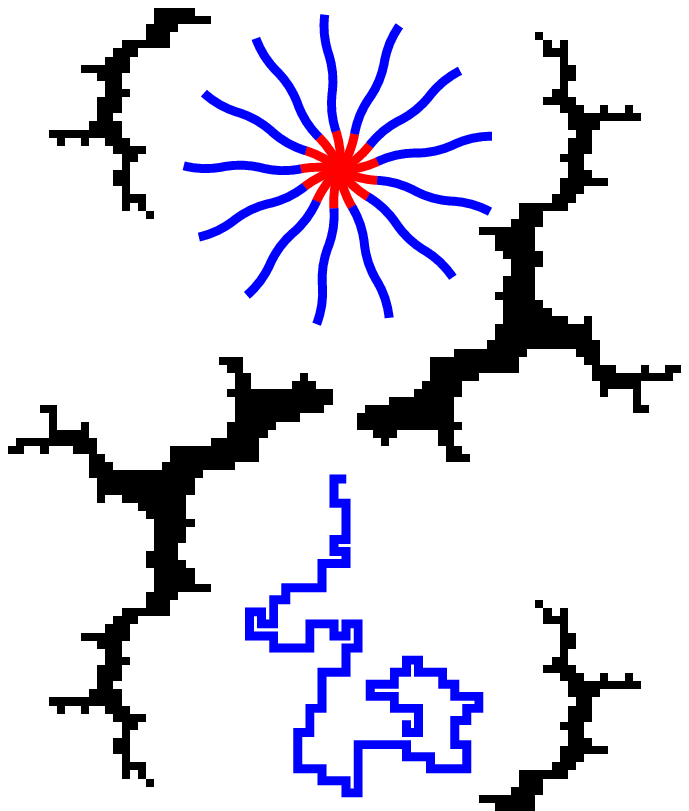,width=1.8in}}
\centerline{\bf a. \hspace{2.5in} b.} \caption{Solution of star and
chain polymers (blue on-line) in a good solvent immersed in a porous
medium (black). We consider the disorder that characterizes pores
(impurities) distribution as uncorrelated ({\bf a}) or correlated
({\bf b}). For the $d$-dimensional system, disorder is called long
range correlated if the impurity-impurity pair correlation function
decays for the large distances $r$ as $g(r)\sim r^{-a}$ with $a< d$.
} \label{hol:fig1}
\end{figure}

One of the generalizations of the above approach extends the theory
to describe polymers of complex structure that form networks of
interconnected polymer chains.\cite{Schaefer92} The intrinsic exponents
that govern the scaling of a polymer network are uniquely defined by
those of its constituents, star-like subunits known as star
polymers\cite{Duplantier89} (see Fig. \ref{hol:fig1}).
The exponents that govern the scaling of star polymers
are universal in that they depend on space dimension $d$ and star
functionality  $f$
only (the number of chains attached to a common center).\footnote{Note, that for $f=1,f=2$ a chain polymer is
recovered.} Currently, star polymers are synthesized with
high functionalities and form well-defined objects with
interesting industrial, technological, and experimental
applications.\cite{stars}

In this paper we attract attention  to another recent development in
the analysis  of the scaling properties of branched
polymers.\cite{Blavatska01,Blavatska06} Our analysis concerns the
impact of structural disorder on the scaling of polymer stars and
chains. For lattice models, where polymers are viewed as
self-avoiding walks (SAW) this type of disorder may be realized by
forbidding the SAW to visit certain lattice sites, which might be
interpreted as  lattice dilution. For real polymers in solvents,
structural disorder may be implemented by filling a porous medium by
a solvent with immersed polymers as shown in Fig.~\ref{hol:fig1}. As
follows from our analysis sketched below, for long-range-correlated
disorder the polymer behaviour displays a new universality,
different from that of a SAW on a regular (undiluted) lattice. A
notable consequence is that star and chain polymers react in  {\em
qualitatively} different manner on the presence of disorder: the
corresponding scaling exponents {\em increase} for the chains and
{\em decrease} for the stars. We discuss to which physical
consequences such a difference in behaviour may lead.

\section{Model and method}\label{hol:sec2}

The starting point of our analysis is the Edwards continuous chain
model, generalized to describe a branched polymer structure, a star
polymer. We describe the conformation of each arm of the star by a path
${\bf r}_a(s)$, parameterized by $0\leq s\leq S_a$, $a=1,2,\ldots,f$
($S_a$, the Gaussian surface of the $a$-th arm is related to the contour
length of the chain), ${\bf r}_a(0)$,
corresponds to the central point. The partition function of the
system is defined by the path integral:\cite{Schaefer92}
\begin{equation} \label{hol:eq1}
{\cal Z} \{S_a\}= \int D [ {\bf r}_1,\ldots, {\bf r}_f ]  \exp
\left[ -\frac{H_f}{k_BT} \right]\prod_{a=2}^f\delta^d({\bf
r}_a(0)-{\bf r}_1(0)).
\end{equation}
Here, the product of $\delta$-functions ensures the star-like
configuration of the set of $f$ polymer chains described by the
Hamiltonian:
\begin{equation} \label{hol:eq2}
\frac{H_f}{k_BT} =\frac{1}{2}\sum_{a=1}^f\int_0^{S_a}{\rm d}\,s
\left(\frac{{\rm d}\,{\bf r}(s)}{{\rm d} s}\right)^2+ \frac{u_0}{4!}
\sum_{a,b=1}^f \int_0^{S_a}{\rm d} s\int_0^{S_b}{\rm d}s'
\delta^d({\bf r}_a(s)-{\bf r}_b(s')).
\end{equation}
The first term in the r.h.s. of (\ref{hol:eq2}) presents chain
connectivity whereas the second term describes an excluded volume
interaction. Instead of introducing structural disorder directly
into Eq. (\ref{hol:eq1}), we make use of its field theoretical
representation. The corresponding derivations are described in
details in Refs.\cite{Schaefer92,Blavatska01}. The relevant steps
read:
\newline ({\bf i}) we map the continuous chain model
(\ref{hol:eq1}) onto the $m=0$ limit of $O(m)$ symmetric field
theory by a familiar Laplace transform\cite{desCloizeaux90} in the
Gaussian surface variables $S_a$ to conjugated chemical potentials
(mass variables) $\mu_a$. In this procedure, the product of
$\delta$-functions in (\ref{hol:eq1}) is represented by a composite
operator of a product of $f$ $m$-component fields
\begin{equation}\label{hol:eq3}
 \sum_{\{k\}} \sum_{i_1,\ldots,i_f=1}^m
N^{i_1,\ldots,i_f}\phi^{i_1}_{k_1} \ldots\phi^{i_f}_{k_f}.
\end{equation}
Here, $N^{i_1,\ldots,i_f}$ is a traceless tensor, $\phi^{i}$ is an
$i$-th component of the $m$-vector field $\vec{\phi}$ and in the sum
over wave vectors $\{k\}$  is restricted by momentum conservation;

\noindent  ({\bf ii}) we introduce quenched random-temperature-like
disorder shifting $\mu_a \rightarrow \mu_a + \delta \mu_a(x)$ by
random variables $\delta \mu_a(x)$. These have zero mean and
correlations that decay at large distances as a power
law:\cite{Weinrib83}
\begin{equation}\label{hol:eq4}
 \langle\delta\mu_a(x)\delta\mu_a(y)\rangle \sim
|x-y|^{-a}.
\end{equation}
Here, $\langle \dots \rangle$ stands for the configurational average
over spatially homogeneous and isotropic disorder and the exponent
$a$ governs the correlation decay. As seen below, this leads to long
range correlated disorder effects for $a<d$;

\noindent  ({\bf iii}) to perform the configurational average of the
free energy, we make use of the replica method resulting in field
theoretical Lagrangean of two couplings $u_0$, $w_0$:
\begin{eqnarray}
{\cal L}&=&\frac{1}{2}\sum_{\alpha=1}^n
\sum_{k}(\mu_0^2+k^2)(\vec{\phi}^{\alpha}_k)^2 + \frac{u_0}{4!}
\sum_{\alpha=1}^n \sum_{\{k\}} \left( \vec{\phi}_{k_1}^{\alpha}
\cdot \vec{\phi}_{k_2}^{\alpha} \right) \left(
\vec{\phi}_{k_3}^{\alpha} \cdot \vec{\phi}_{k_4}^{\alpha} \right) +
\nonumber\\
 &&\frac{w_0}{4!}\sum_{\alpha , \beta=1}^n \sum_{\{k\}}
  |k_1{+}k_2|^{a-d}
\left(\vec{\phi}_{k_1}^{\alpha} \cdot
\vec{\phi}_{k_2}^{\alpha}\right) \left(\vec{\phi}_{k_3}^{\beta}
\cdot \vec{\phi}_{k_4}^{\beta}\right).\label{hol:eq5}
\end{eqnarray}
Note that the evaluation of the theory (\ref{hol:eq5}) involves a
simultaneous polymer ($m=0$) and replica ($n=0$) limit. It is this
anticipated double limit that allows us\cite{Blavatska01} to write
the Lagrangean in terms of two coupling only. A third coupling
appears\cite{Weinrib83} for $m\neq 0$.

\begin{figure}
\psfig{file=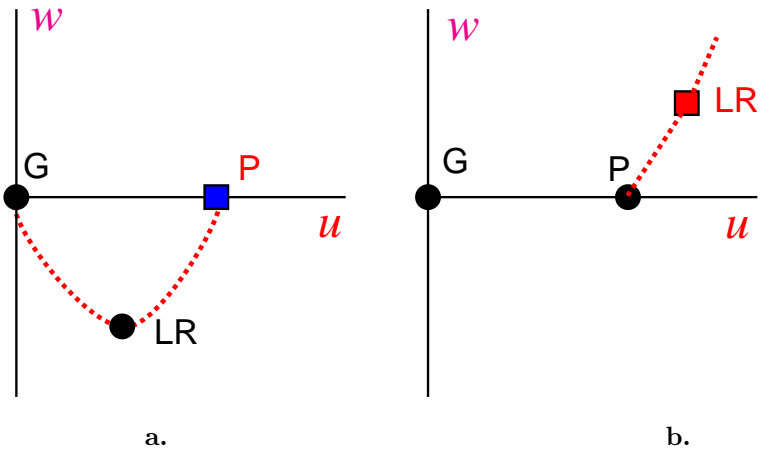,width=4.1in} \centerline{\bf a.
\hspace{2.5in} b.} \caption{Fixed points G (gaussian), P (pure) and
LR (long range) of the RG flow in the plane of the two couplings
$u,w$. The stable FP is shown by a square.({\bf a}) uncorrelated
disorder $a>d$: P ($u\neq 0, w=0$)  is stable. ({\bf b}) correlated
disorder $a<d$: LR ($u\neq 0, w\neq 0$) is stable. Crossover occurs
at $a=d$ (c.f. Fig. \ref{hol:fig1}). } \label{hol:fig2}
\end{figure}

We apply the field theoretical renormalization group (RG) approach
to extract the universal content of (\ref{hol:eq5}). In this
approach, the change of the couplings $u, w$ under renormalization
defines a flow in parametric space, governed by corresponding
$\beta$-functions $\beta_{u}(u,w)$, $\beta_{w}(u,w)$. The fixed
points (FPs) $u^{*},w^{* }$ of this flow are the solutions to the
system of equations: $ \beta_u(u^{*},w^{*})=0,\,
\beta_w(u^{*},w^{*})=0$. If a stable FP exists and is
accessible, it determines the scaling behaviour of the
polymer system. In particular, for a single polymer star of
$f$ arms of equal length $N$ in a good solvent the partition sum
(number of possible configurations) scales as
\begin{equation} \label{hol:eq6}
Z_{N,f} \sim {\rm e}^{\mu Nf}N^{\gamma_f-1}, \hspace{2em} N \to
\infty ,
\end{equation}
with a non-universal fugacity ${\rm e}^{\mu}$ and the universal star
exponent $\gamma_f$.\cite{Duplantier89} The latter is uniquely
defined by the stable FP value of the anomalous dimension associated
with the composite operator (\ref{hol:eq3}). We make use of two
complementary perturbation theory expansions to calculate
coordinates of the FPs and values of the exponents. In a first
approximation we apply an expansion in\cite{Weinrib83}
$\varepsilon=4-d$ and $\delta=4-a$ which allows for a qualitative
description of the phenomena. In a further approach we apply
perturbation theory in the renormalized couplings $u$ and $w$
evaluated at fixed dimension $d=3$ for a series of fixed values of
the correlation parameter $a$.\cite{Prudnikov} In the latter case we
proceed within a two-loop approximation refining the analysis by a
resummation of the divergent RG series (see
Ref.\cite{Blavatska01,Blavatska05,Blavatska06} for details). The FP
picture that arises from our calculations is shown qualitatively in
Fig. \ref{hol:fig2}.
Both calculation schemes display a range $a_{\rm lower}(d)<a<a_{\rm
upper}(d)$, of values for $a$ where the long-range-correlated FP
(LR, $u\neq 0$, $w\neq 0$) is stable and governs polymer scaling.
For $a<a_{\rm lower}$, no stable FP is found. This has been
interpreted\cite{Blavatska01} as a collapse of a polymer coil for
strongly correlated disorder. For $a>a_{\rm upper}$ the pure FP (P,
$u\neq 0$, $w=0$) is stable and polymer scaling is not perturbed by
disorder. As far as power counting implies that the $w$-term in
(\ref{hol:eq5}) is irrelevant in the RG sense for $a\geq d$ it is
natural to identify $a_{\rm upper}=d$. Nonetheless, in first order
approximation of the $\varepsilon,\delta$-expansion one finds that
the LR FP is stable in the unphysical region $d < a < 2 +d/2$. Our
two-loop calculations at fixed $d,a$ for $d=3$ however result in
$a_{\rm upper}=3=d$, $a_{\rm lower}=2.2$ allowing for direct
physical interpretations.

\section{Scaling exponents}\label{hol:sec3}

Qualitatively, the impact of disorder can be seen already from the
first order $\varepsilon,\delta$ results. Comparing the
$\gamma_f$ exponent (\ref{hol:eq6}) for the cases when the
structural disorder is absent (or is short-range-correlated, Fig.
\ref{hol:fig1}a),\cite{Duplantier89} $\gamma_f^{(0)}=1- \varepsilon
f(f-3)/16$, and when it is long-range-correlated (Fig.
\ref{hol:fig1}b)\cite{Blavatska06}, $\gamma_f^{(\delta)}=1-\delta
f(f-3)/8$ we find:
\begin{equation} \label{hol:eq7}
\Delta \gamma_f \equiv \gamma_f^{(\delta)} - \gamma_f^{(0)} =
\frac{f(f-3)}{16} (\varepsilon - 2 \delta), \hspace{2em}
\varepsilon/2 < \delta < \varepsilon.
\end{equation}
\begin{figure}
\begin{center}
\psfig{file=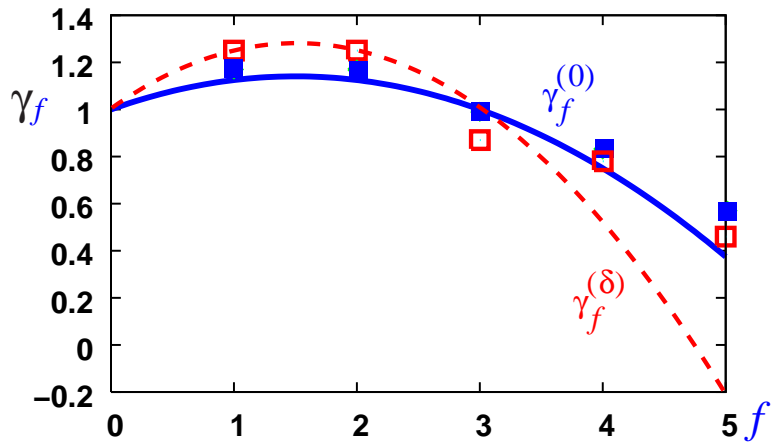,width=4.1in}
\end{center}
\caption{Exponent $\gamma_f$ of 3D polymer stars with $f$ arms in a
good solvent. Solid lines and filled squares indicate uncorrelated
disorder, dashed lines $\delta=1.1$ and open squares $a=2.9$
correlated disorder; lines: first order
$\varepsilon,\delta$-results; open, resp. filled, squares: second,
resp. third, order resummed results for fixed  $d,a$.}
\label{hol:fig3}
\end{figure}
As one can see from this estimate, the  exponent difference changes
sign at $f_{\pm}=3$. The two-loop calculations slightly shift this
result towards
 $f_{\pm}\leq 3$ while otherwise confirming the overall picture
with more accurate numerical values for the
exponents.\cite{Blavatska05,Blavatska06} In Fig. \ref{hol:fig3} we
plot the first order $\varepsilon,\delta$-curves for $\gamma_f$
together with the resummed two-loop estimates for $d=3$.
A complete account of the numerical values of the exponents is given in
Ref.\cite{Blavatska06} A prominent feature that follows from these
results is that the effect of long-range-correlated disorder
on polymer chains and polymer stars
 is qualitatively different.
Whereas correlated disorder leads to an {\em increase} of $\gamma_f$
for chain polymers (i.e. for $f=1$, $f=2$), the same type of
disorder {\em decreases} the $\gamma_f$ exponent for the proper star
polymers ($f\geq 3$). Below we discuss some possible consequences of
this difference. For $f>2$ we also observe that  $\gamma_f$
decreases monotonically as function of $f$ for any valid value of
$a$ in the same way as it does for polymers in a good solvent:
$\gamma_{f_1}^{(a)} > \gamma_{f_2}^{( a)}$  for $2<f_1<f_2$.
Furthermore, within the accuracy the data confirms a convexity from
below of this function for valid values of $a$ as well as for the
$\varepsilon,\delta$ expansion. The latter property ensures that
polymer stars remain mutually repulsive in correlated
disorder\cite{copolymer,Duplantier91}.

\section{Conclusions and outlook}\label{hol:sec4}

It has been recognized by now that star exponents come into play for
the qualitative description and quantitative of different phenomena,
where statistics of branched self avoiding and random walks is
involved. The examples of such phenomena include short-range
interaction of branched polymers in a good solvent),\cite{interact}
diffusion-controlled reactions in the presence of
polymers,\cite{dla} and, more generally, they are part of a
multifractal description of diffusion limited growth in a Laplacian
field.\cite{MF} Recently, star exponents have been used to estimate
the thermal denaturation transition of DNA.\cite{Carlon02} Our
analysis opens a way to consider the impact of long-range-correlated
disorder on the above phenomena.

A somewhat surprising  effect that results from our calculations
concerns the static separation in a solution of diluted chains and
star polymers of equal molecular weight inside a porous medium.
Following the estimates of the star exponents we predict that in a
correlated medium star polymers will exert a higher osmotic pressure
than chain polymers and in general higher branched star polymers
will be expelled more strongly from the correlated porous medium. On
the opposite, polymer chains will prefer a stronger correlated
medium to a less or uncorrelated medium of the same
density.\cite{Blavatska06}

A generalization of our approach to the case of star polymers
built from chains of different species\cite{copolymer} will be
presented elsewhere.\cite{Blavatska07}

\section*{Acknowledgments}
We acknowledge support by the  EU Programme  ``Marie Curie
International Incoming Fellowship" (V.B.) and Austrian Fonds zur
F\"orderung der wi\-ssen\-schaft\-li\-chen Forschung under Project P
19583 - N20 (Yu.H.).

\end{document}